\documentclass[twocolumn,showpacs,pra,superscriptaddress,longbibliography]{revtex4}
\usepackage{times}
\usepackage{amsmath}
\usepackage{amssymb}
\usepackage{graphicx}
\usepackage[unicode]{hyperref}
\hypersetup{
   unicode=true,          
   plainpages=false,
   colorlinks=true,       
   citecolor=blue,        
}
\usepackage[caption=false,position=top,singlelinecheck=off,justification=raggedright]{subfig}
\usepackage{color}
\usepackage{pst-node}

\def \b1{{\bf 1}}

\newcommand{\bea}{\begin{eqnarray}}
\newcommand{\eea}{\end{eqnarray}}
\newcommand{\beq}{\begin{equation}}
\newcommand{\eeq}{\end{equation}}

\begin{document}

\title{From a continuous to a discrete time crystal in a dissipative atom-cavity system}

\author{Hans Ke{\ss}ler}
\affiliation{Zentrum f\"ur Optische Quantentechnologien and Institut f\"ur Laser-Physik, 
Universit\"at Hamburg, 22761 Hamburg, Germany}
\affiliation{The Hamburg Center for Ultrafast Imaging, Luruper Chaussee 149, Hamburg 22761, Germany}

\author{Jayson G. Cosme}
\affiliation{Zentrum f\"ur Optische Quantentechnologien and Institut f\"ur Laser-Physik, 
Universit\"at Hamburg, 22761 Hamburg, Germany}
\affiliation{The Hamburg Center for Ultrafast Imaging, Luruper Chaussee 149, Hamburg 22761, Germany}

\author{Christoph Georges}
\affiliation{Zentrum f\"ur Optische Quantentechnologien and Institut f\"ur Laser-Physik, 
Universit\"at Hamburg, 22761 Hamburg, Germany}

\author{Ludwig Mathey}
\affiliation{Zentrum f\"ur Optische Quantentechnologien and Institut f\"ur Laser-Physik, 
Universit\"at Hamburg, 22761 Hamburg, Germany}
\affiliation{The Hamburg Center for Ultrafast Imaging, Luruper Chaussee 149, Hamburg 22761, Germany}

\author{Andreas Hemmerich}
\affiliation{Zentrum f\"ur Optische Quantentechnologien and Institut f\"ur Laser-Physik, 
Universit\"at Hamburg, 22761 Hamburg, Germany}
\affiliation{The Hamburg Center for Ultrafast Imaging, Luruper Chaussee 149, Hamburg 22761, Germany}


\date{\today}
\begin{abstract}
We propose the dynamical stabilization of a nonequilibrium order in a driven dissipative system comprised an atomic Bose-Einstein condensate inside a high finesse optical cavity, pumped with an optical standing wave operating in the regime of anomalous dispersion. When the amplitude of the pump field is modulated close to twice the characteristic limit-cycle frequency of the unmodulated system, a stable subharmonic response is found. The dynamical phase diagram shows that this subharmonic response occurs in a region expanded with respect to that where stable limit-cycle dynamics occurs for the unmodulated system. In turning on the modulation we tune the atom-cavity system from a continuous to a discrete time crystal.
\end{abstract}

\maketitle

\section{Introduction}
A time crystal is a nonequilibrium phase of matter that spontaneously breaks time translation symmetry \cite{Wilczek2012}. Following a series of no-go theorems that questioned the possibility of observing time crystals as a ground state in equilibrium systems \cite{Bruno2013, Noz2013, Watanabe2015}, the attention has shifted towards far-from-equilibrium scenarios, namely periodically driven isolated \cite{Sacha2015, Sacha2018, Else2019, Khemani2019, Else2016, Yao2017, Choi2017, Zhang2017,Ho2017, Else2017, Rovny2018, Smits2018, Giergiel2018,Giergiel2019, Pizzi2019, Autti2018}  and dissipative \cite{Gong2018, Gambetta2018, Iemini2018, Tucker2018, Heugel2019, Lazarides2019, Zhu2019,Lledo2019, Buca2019, Kessler2019, Cosme2019, Seibold2020, Yao2020,Lledo2020} systems. A discrete time crystal (DTC) emerges when the system spontaneously breaks the discrete time translation symmetry imposed by an external drive with period $T$, which manifests as a subharmonic response of observables $\langle O(t) \rangle =\langle O(t+nT) \rangle$ with $n \in \{2,3,\dots \}$. On the other hand, a continuous time crystal (CTC) forms when a system exhibits a robust characteristic frequency without periodic driving. An example of a CTC is the limit cycle phase predicted in an atom-cavity system \cite{Keeling2010, Piazza2015, Kessler2019}. Here, we propose a stabilization of this dynamical phase by periodic driving, which effectively leads to a conversion of a continuous time crystal into a discrete time crystal.

\begin{figure}[!htb]
\centering
\includegraphics[width=1.0\columnwidth]{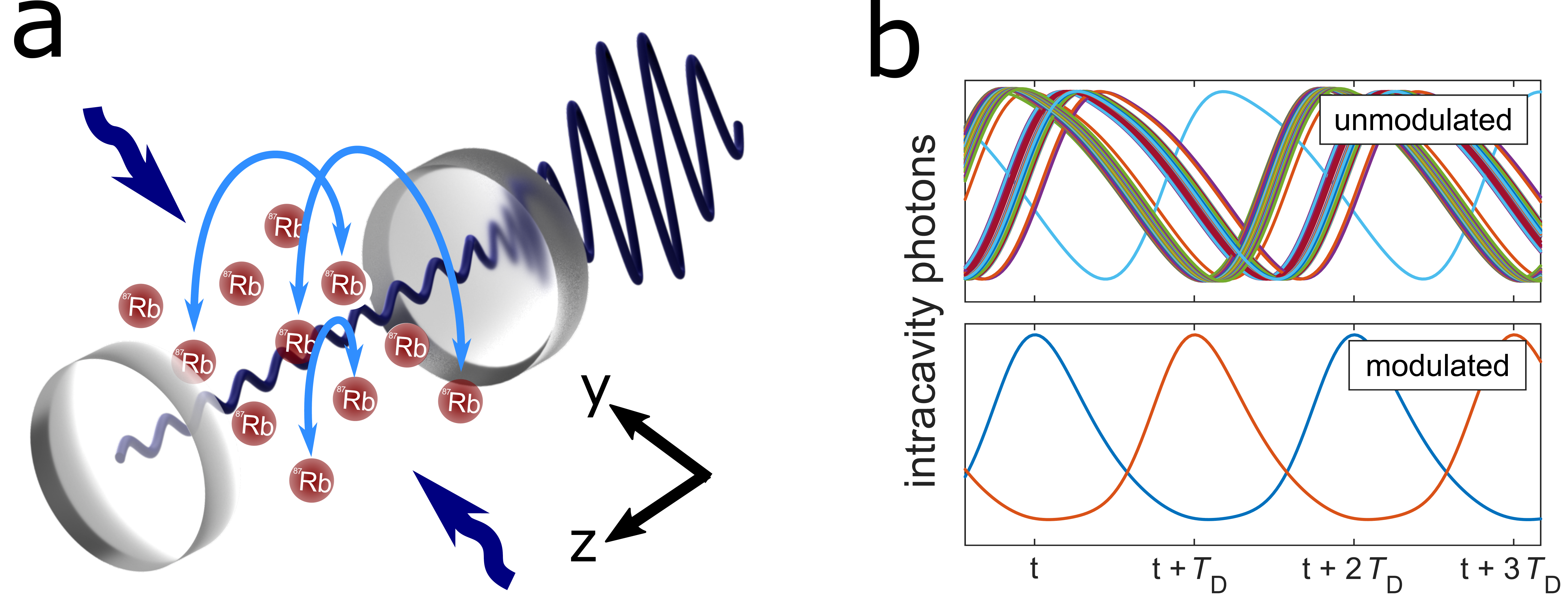}
\caption{(a) A Bose-Einstein condensate is prepared inside a high finesse optical cavity. Oriented perpendicularly to the cavity axis are two counterpropagating pump beams that provide a repulsive standing wave light-shift potential due to its frequencies being blue-detuned with respect to the atomic resonance. (b) Dynamics of the intracavity photon number using different initial conditions. Upper panel: In the absence of temporal modulation of the pump field, the system enters a limit cycle or continuous time crystalline phase for strong pump intensities. Different colors indicate different noise-induced initial conditions. Lower panel: For near-resonant modulation of the pump field at driving period $T_\mathrm{D}$, the limit-cycle phase is stabilized and transformed into a period-doubled discrete time crystal. The colors indicate the two possible noise-induced phase positions relative to the modulation.}
\label{fig:1} 
\end{figure}

In this work, we explore the fundamental effect of parametric driving in a system of an atomic Bose-Einstein condensate (BEC) inside a high finesse optical cavity and pumped by an optical standing wave oriented perpendicularly with respect to the cavity axis, as depicted in Fig.~\ref{fig:1}(a) \cite{Ritsch2013}. The dissipative cavity not only provides a cold bath that prevents heating, but it also induces an all-to-all coupling between the atoms. This infinite range interaction is an important aspect of mean-field time crystals \cite{Gong2018, Zhu2019, Cosme2019}. We are interested in the case when photons are pumped by a repulsive standing wave. This regime can be experimentally realized using positive or blue detuning of the pump with respect to the relevant atomic resonance \cite{Zupancic2019}. In addition to the self-organized superradiant phase found in its red-detuned counterpart \cite{Domokos2002,  Baumann2010, Klinder2015}, the less explored regime of a blue-detuned pump field also hosts nonequilibrium phases, which provide limit-cycle \cite{Keeling2010, Piazza2015, Kessler2019,Lin2019} or chaotic \cite{Zupancic2019} dynamics. This results from the positive sign of the light-shift associated with the pump and intra-cavity light fields, which acts to localize the atoms at intensity nodes, thus counteracting the formation of a stationary intra-cavity optical lattice. Here, we focus on a cavity operating in the recoil-resolved regime \cite{Klinder2016}, which is essential for a stable limit-cycle phase, as it allows the cavity to evolve on the same timescale as the atomic motion. The sub-recoil nature of the cavity limits the number of accessible momentum modes, and hence, it is an essential prerequisite for staying away from chaotic or heating dynamics, which is shifted to stronger driving regimes. 

\begin{figure*}[!htp]
\centering
\includegraphics[width=2.0\columnwidth]{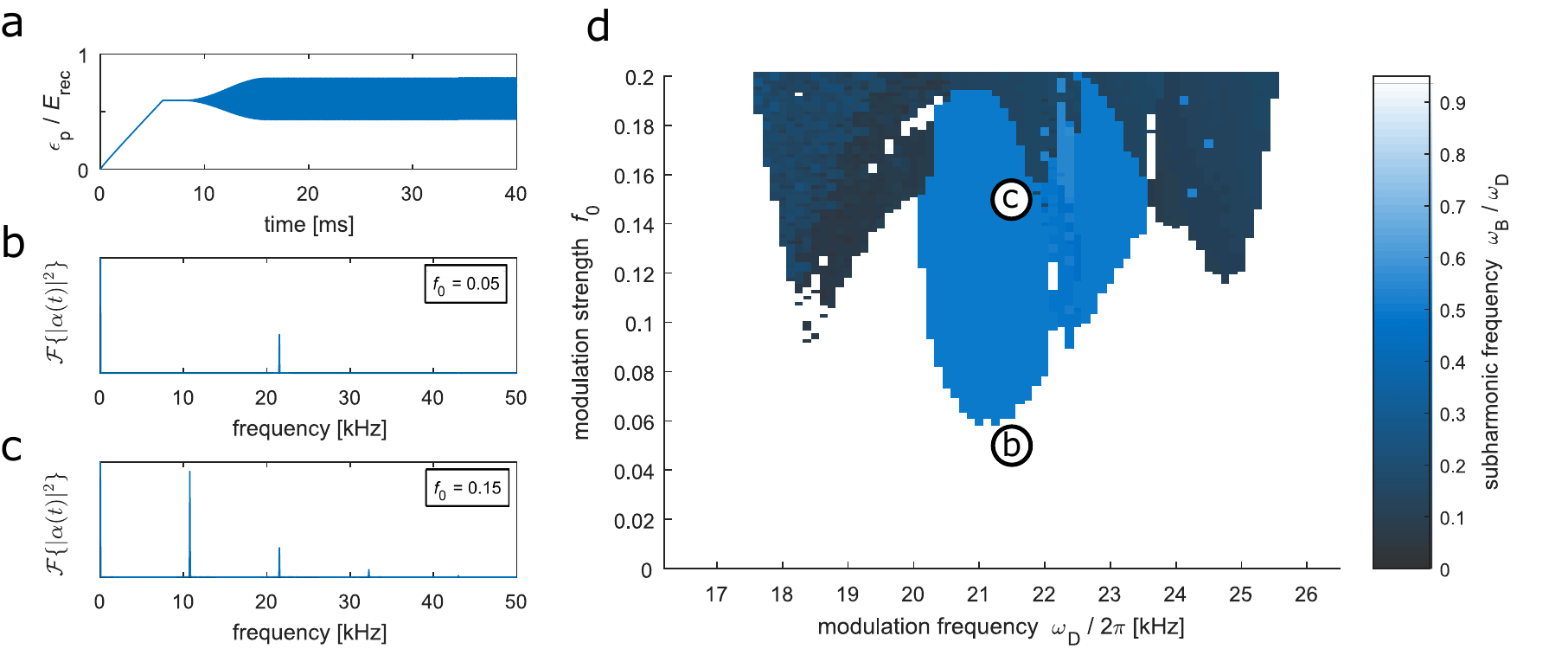}
\caption{(a) Exemplary driving protocol for the pump. (b-c) Typical Fourier spectra of the dynamics of the cavity mode occupation for $\omega_{\mathrm{D}}/2\pi=21.5~\mathrm{kHz}$ with (b) $f_0=0.05$ and (c) $f_0=0.15$. (d) Dependence of the subharmonic frequency on $f_0$ and $\omega_{\mathrm{D}}$ for fixed $\delta_{\mathrm{eff}}/2\pi=-11.5~\mathrm{kHz}$ and $\epsilon_0=0.6~E_\mathrm{rec}$. The white disks with ``b" and ``c" show the locations corresponding to the spectra shown in (b) and (c).}
\label{fig:2} 
\end{figure*} 

We propose to stabilize the CTC order in the atom-cavity system by periodically driving the amplitude of the pump field at frequencies $\omega_{\mathrm{D}}$ close to the primary parametric resonance of the limit-cycle phase, $\omega_{\mathrm{D}} = 2\omega_{\mathrm{LC}}$. The response of the system for near-resonant driving is characterized by a perfect period-doubling dynamics of the cavity mode occupation, which suggests that the dynamical phase turns into a DTC phase as schematically shown in Fig.~\ref{fig:1}(b). In doing so, the rigidity of the limit cycle is enhanced as it occupies an increased area in the relevant parameter space. We emphasize that our findings suggest the unique possibility of stabilizing an already dynamical order, shown to qualify as a CTC \cite{Kessler2019}, by appropriate driving. This is in contrast to previous studies focusing on driving-induced stabilization of static orders \cite{Georges2018, Cosme2018}.

\section{Atom-cavity system}\label{sec:system}

In Fig.~\ref{fig:1}(a), we present a schematic diagram of the atom-cavity system consisting of a Bose-Einstein condensate (BEC) trapped inside a high finesse optical resonator and pumped in the transverse direction by an optical standing wave, where the pump and cavity axes are along the $y$ and $z$ direction, respectively. The pump is characterized by its wavelength $\lambda$ and wavenumber $k=2\pi/\lambda$.  The strength of the pump is parametrized in terms of the potential depth of the lattice formed in units of the recoil energy $E_{\mathrm{rec}}=\hbar\omega_{\mathrm{rec}}$, where $\omega_{\mathrm{rec}}=\hbar k^2/2m$ and $m$ is the atomic mass. The dissipative cavity has a field decay rate $\kappa$.

We consider the two circularly polarized intracavity modes and the linearly polarized transverse pump field \cite{Kessler2019}. The corresponding equations of motion for the BEC mode $\Psi(y,z)$ and the two cavity modes ${\alpha}_{\pm}$ with polarization components $\zeta_{\pm}$ are
\begin{align}\label{eq:eom}
i\hbar\frac{\partial {\Psi}(y,z)}{\partial t} &= \left(-\frac{\hbar^2}{2m}\nabla^2 + U_{\mathrm{dip}}(y,z)\right) {\Psi}(y,z) \\ \nonumber
i\frac{\partial {\alpha}_{\pm}}{\partial t} &= \left(-\delta_c + U_{\pm}\mathcal{B} -i\kappa\right){\alpha}_{\pm} + \frac{\alpha_{\mathrm{p}}(t)}{\sqrt{2}} U_{\pm}\Phi + i\xi.
\end{align}
The detuning between the pump frequency and the empty cavity resonance is $\delta_c$. The bunching parameter is $\mathcal{B}=\langle \mathrm{cos}^2(kz) \rangle$ and the density wave order parameter corresponding to the checkerboard ordering is $\Phi=\langle \mathrm{cos}(kz)\mathrm{cos}(ky) \rangle$. We neglect the effects of short-range atomic interactions and of the trapping potentials. 
The dipole potential $U_{\mathrm{dip}}$ due to the pump and cavity fields,  $g(y)=\mathrm{cos}(ky)$ and $f(z)=\mathrm{cos}(kz)$, respectively, is 
\begin{align}
&U_{\mathrm{dip}}(y,z)=\hbar U_0\biggl[ f(z)^2\left(\zeta^2_{-}|\alpha_{-}|^2 + \zeta^2_{+}|\alpha_{+}|^2\right) + \frac{|\alpha_{\mathrm{p}}(t) g(y)|^2 }{2}  \\ \nonumber
& + \frac{\alpha_{\mathrm{p}}(t)}{\sqrt{2}}f(z)g(y)\left( \zeta^2_{-}(\alpha_{-}+\alpha^{*}_{-}) +\zeta^2_{+}(\alpha_{+}+\alpha^{*}_{+}) \right) \biggr],
\end{align} 
where $U_0$ is the light shift per intracavity photon.
Note that in Eq.~\eqref{eq:eom}, we introduced polarization-dependent light shift $U_{\pm}=U_0 \zeta_{\pm}^2$. 
We are interested in the response from periodic driving of the amplitude of the pump field
\begin{equation}
\alpha_{\mathrm{p}}(t) = \alpha_{0}\left(1 + f_0 \mathrm{cos}(\omega_{\mathrm{D}}t) \right).
\end{equation}
The driving period is $T_{\mathrm{D}}=2\pi/\omega_{\mathrm{D}}$. 
In units of the recoil energy, the time-dependent pump strength is $\epsilon_{\mathrm{p}} = {|{\alpha_{\mathrm{p}}}|^2 U_0}/{2\omega_{\mathrm{rec}}}$. It is insightful to keep track of the zero-frequency (pump frequency in the lab frame) component of the pump intensity given by $\epsilon_0 = {|{\alpha_{\mathrm{0}}}|^2 U_0}/{2\omega_{\mathrm{rec}}}$.  Note that in Eq.~\eqref{eq:eom}, we include the stochastic noise term $\xi(t)$ satisfying $\langle \xi^*(t)\xi(t') \rangle = \kappa \delta(t-t')$ in the time evolution of the cavity mode when obtaining results beyond the mean-field approximation \cite{Cosme2018,Kessler2019,Cosme2019,Georges2018}.

In the following simulations, we use realistic parameters based on Ref.~\cite{Klinder2016}, which are $\omega_{\mathrm{rec}}=2\pi \times 3.872~\mathrm{kHz}$, $\kappa=2\pi \times 4.50~\mathrm{kHz}$, $U_0= 2\pi \times 1.14~\mathrm{Hz}$, $\zeta^2_{+}=0.68$, $\zeta^2_{-}=0.32$, and  $N_a = 60\times 10^3$ the number of atoms. We solve the set of coupled equations in Eq.~\eqref{eq:eom} by expanding the BEC wavefunction in the plane-wave basis ${\Psi}(y,z) = \sum_{n,m}{\phi}_{n,m}\mathrm{e}^{i n k y}\mathrm{e}^{i m k z}$, where ${\phi}_{n,m}$ is a single-particle momentum mode with momentum $(n,m)\hbar k$. We find that $\{n,m\}\in [-6,6]~\hbar k$ yields numerically convergent results. 

\begin{figure*}[!htbp]
\centering
\includegraphics[scale=1.0]{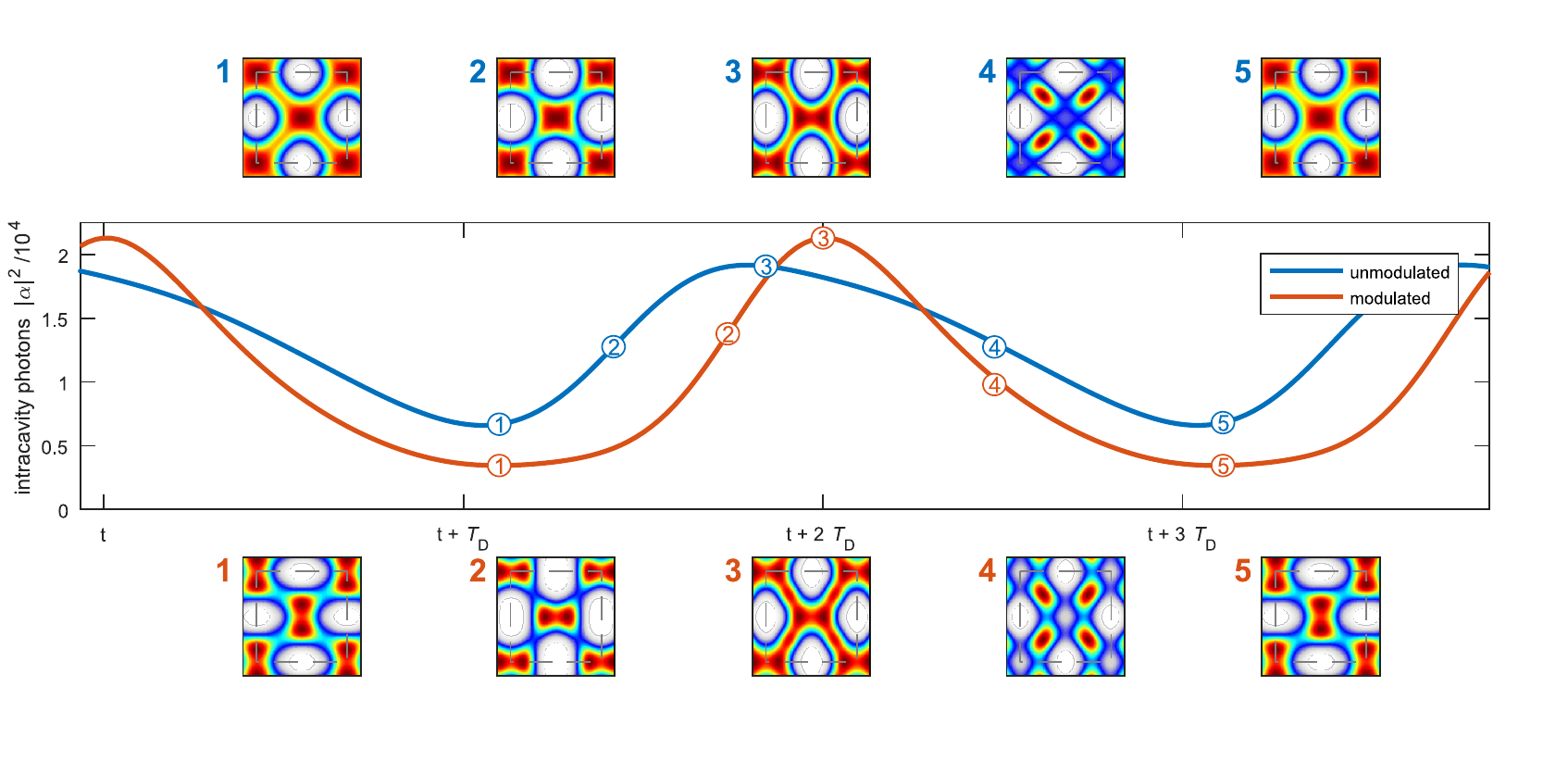}
\caption{Comparison between the response from unmodulated and modulated pump fields. For the unmodulated case, the pump strength is constant at $\epsilon_0=0.8~E_{\mathrm{rec}}$. For the modulated case, the modulation parameters are $f_0=0.15$ and $\omega_{\mathrm{D}}/2\pi = 21.2~\mathrm{kHz}$ with $\epsilon_0=0.6~E_{\mathrm{rec}}$. In both cases, the effective detuning is fixed at $\delta_{\mathrm{eff}}/2\pi=-11.5~\mathrm{kHz}$. Snapshots of the single-particle distributions of the atomic ensemble taken at various instances of time, as marked in the middle panel, are shown in the top and bottom panels for the unmodulated and modulated cases, respectively. The gray dashed squares indicate the unit cell.}
\label{fig:3} 
\end{figure*} 
\section{Response from periodic driving}\label{sec:resp}

For the driving protocol considered here, the pump strength $\epsilon_{\mathrm{p}}$ is first linearly ramped from zero to the desired value $\epsilon_{\mathrm{0}}$ at a rate of $0.1~E_{\mathrm{rec}}/\mathrm{ms}$. Then, the pump strength is held at $\epsilon_0$ for 2 ms and the modulation strength $f_0$ is slowly increased over 10 ms. Finally, the pump field is continuously modulated until the entire protocol reaches 40 ms. This allows the system to approach its long-time behavior. An exemplary time series of the pump strength is shown in Fig.~\ref{fig:2}(a).

We focus on the dynamics of the $\sigma^+$-polarized cavity mode occupation $|\alpha_{+}|^2 \equiv |\alpha|^2$ as the $\sigma^-$-mode shows qualitatively similar dynamics. In particular, we obtain the Fourier spectra of $|\alpha(t)|^2$ over the final 10 ms of continuous driving, as exemplified in Figs.~\ref{fig:2}(b) and \ref{fig:2}(c). In order to characterize the response of the system from the periodic modulation of the pump, we vary the modulation strength and frequency for fixed values of the effective detuning $\delta_{\mathrm{eff}} \equiv \delta_c-(1/2)N_aU_0$ and the zero-frequency component of the pump strength $\epsilon_0$. Similar to the driven red-detuned pump \cite{Chitra2015, Molignini2018, Cosme2018,Georges2020}, we also find here the appearance of resonance lobes or ``Arnold tongues'' for frequencies associated with parametric resonances. However, a unique feature of the blue-detuned pump is the appearance of a resonance lobe corresponding to a parametric resonance of the limit-cycle phase. 

As has been discussed in Ref.~\cite{Kessler2019}, in the absence of periodic modulation, a CTC phase defined by a limit-cycle frequency $\omega_{\mathrm{LC}}$ emerges for sufficiently strong $\epsilon_0$. In general, the order of magnitude of the limit-cycle frequency $\omega_{\mathrm{LC}}$ is set by the recoil frequency $\omega_{\mathrm{rec}}$ \cite{Keeling2010,Piazza2015,Kessler2019}. Moreover, the limit-cycle frequency $\omega_{\mathrm{LC}}$ shows a weak dependence on $\delta_{\mathrm{eff}}$, $\epsilon_0$, and $\kappa$ \cite{Piazza2015,Kessler2019}. 
In the absence of modulation $f_0=0$, the system is in the density wave (DW) phase for the specific choice of $\delta_{\mathrm{eff}}=-11.5~\mathrm{kHz}$ and $\epsilon_0=0.6~E_{\mathrm{rec}}$ in Fig.~\ref{fig:2}. However, there is a nearby limit-cycle phase at a critical pump strength $\epsilon_0\approx 0.68~E_{\mathrm{rec}}$ with frequency $\omega_{\mathrm{LC}}/2\pi=10.55~\mathrm{kHz}$.

The primary or first-order parametric resonance around the driving frequency $\omega_{\mathrm{D}}=2\omega_{\mathrm{LC}}$ is explored in Fig.~\ref{fig:2}(d).
In this figure, we plot the dominant subharmonic frequency $\omega_{\mathrm{B}}$ defined as 
\begin{equation}
n_\alpha(\omega_{\mathrm{B}})=\max\{n_\alpha(\omega) | \omega<\omega_{\mathrm{D}})\},
\end{equation}
versus the driving frequency $\omega_{\mathrm{D}}$ and the modulation strength $f_0$, where $n_\alpha(\omega)\equiv\mathcal{F}\{ |\alpha(t)|^2 \}$ is the Fourier transform of the cavity mode occupation $|\alpha(t)|^2$. The white region in Fig.~\ref{fig:2}(d) corresponds to the regime with regular response defined by coherent oscillations of the cavity occupation at the modulation frequency, as shown in the singular peak at $\omega_{\mathrm{D}}$ in Fig.~\ref{fig:2}(b). A more exciting physical behavior is found in the blue colored areas in Fig.~\ref{fig:2}(d). Specifically, in the light blue region centered around 21 kHz, the system exhibits perfect period-doubling dynamics $\omega_{\mathrm{B}}=\omega_{\mathrm{D}}/2$ of the cavity mode occupation, as exemplified in the sharp peak at $10.75~\mathrm{kHz}$ in Fig.~\ref{fig:2}(c). This dynamical phase can be classified as a period-doubled DTC order and the robustness of its subharmonic response will be explored later. In the dark blue area in Fig.~\ref{fig:2}(d), the dynamics becomes irregular with multiple side peaks appearing in the Fourier spectra of the cavity dynamics, which suggests that the system is close to the chaotic regime \cite{Kessler2019}.

\begin{figure*}[!htpb]
\centering
\includegraphics[width=2.0\columnwidth]{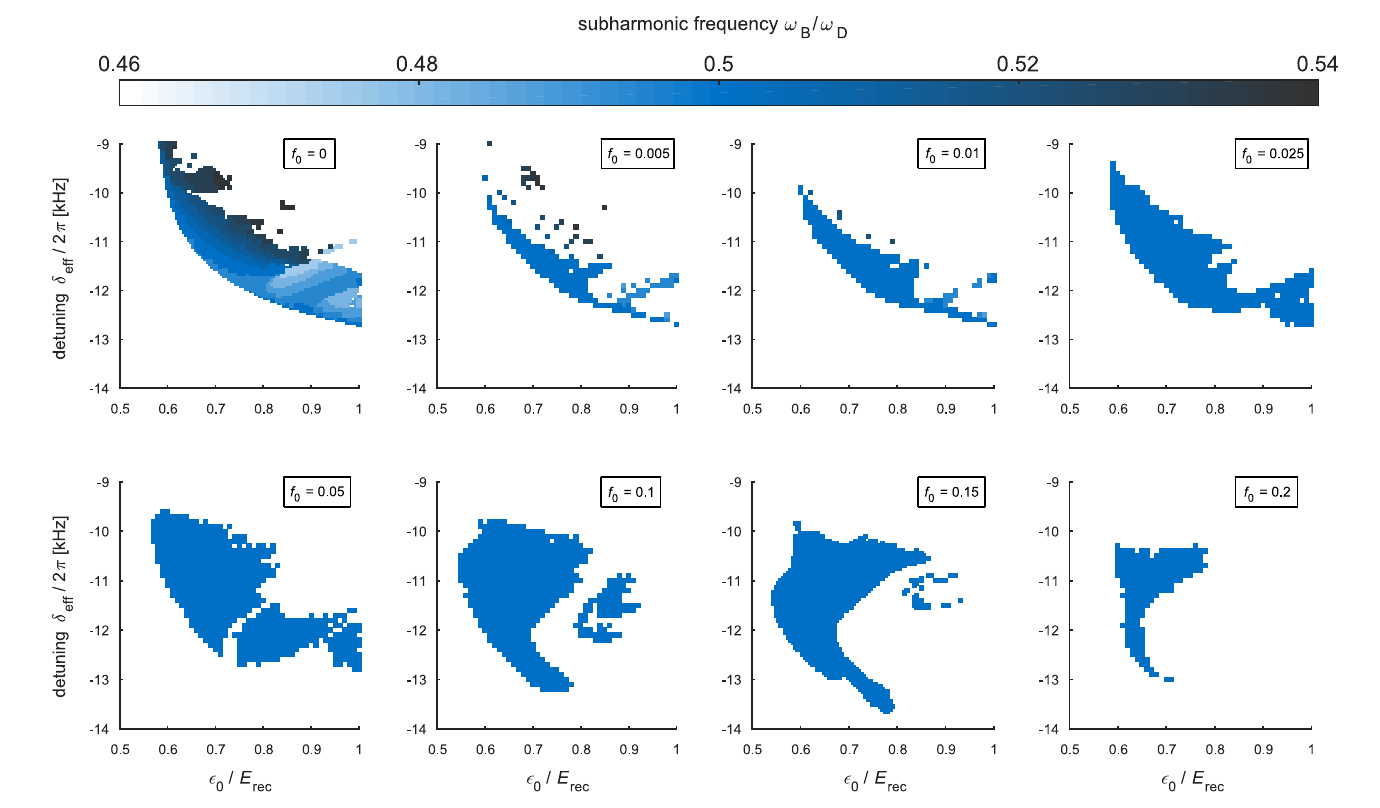}
\caption{Gallery of dynamical phase diagrams for different modulation strengths and fixed modulation frequency $\omega_{\mathrm{D}}/2\pi = 21.2~\mathrm{kHz}$. Each diagram depicts the dependence of the dominant subharmonic frequency $\omega_{\mathrm{B}}$, rescaled by $\omega_{\mathrm{D}}/2\pi = 21.2~\mathrm{kHz}$, on $\delta_{\mathrm{eff}}$ and  $\epsilon_0$ .}
\label{fig:4} 
\end{figure*}
It is instructive to take a closer look at the dynamics for both the cavity and atomic modes in the CTC and DTC phases, appearing when the pump field is unmodulated and modulated, respectively. To this end, we calculate the single-particle density given by
\begin{align}
\rho(y,z) &\equiv \langle \Psi^{\dagger}(y,z)\Psi(y,z) \rangle \\ \nonumber
&= \sum_{n,m,n',m'} \phi^{\dagger}_{n,m}\phi_{n',m'}e^{i(n-n')ky}e^{i(m-m')kz}.
\end{align}
Snapshots of the single-particle density profile are taken at certain instances of time as shown in Fig.~\ref{fig:3}. In contrast to the checkerboard density wave order in the red-detuned case, a significant fraction of atoms self-organize not only in the antinodes but also along the nodes of the pump-cavity potential. Therefore, aside from oscillations in the cavity mode occupation, another distinguishing feature of the limit cycle phase is the presence of Bragg peaks in the momentum distribution of the atomic ensemble at $\pm 2\hbar k$ along the cavity axis. 
More interestingly, similarities between the snapshots of the single-particle density in the CTC and DTC phases reveal that the same set of spatially symmetry-broken states participates in the dynamics of both time crystalline phases. This suggests that the DTC phase is essentially the limit cycle phase found in the absence of periodic driving. We therefore classify both the CTC and DTC phases studied here as limit cycle phases.
Thus, we discover a dynamical protocol for stabilizing an already dynamical phase in the atom-cavity system. In the following, we further investigate the features of this stabilization mechanism.

\section{Dynamical phase diagram}

\begin{figure*}[!htb]
\centering
\includegraphics[width=1.7\columnwidth]{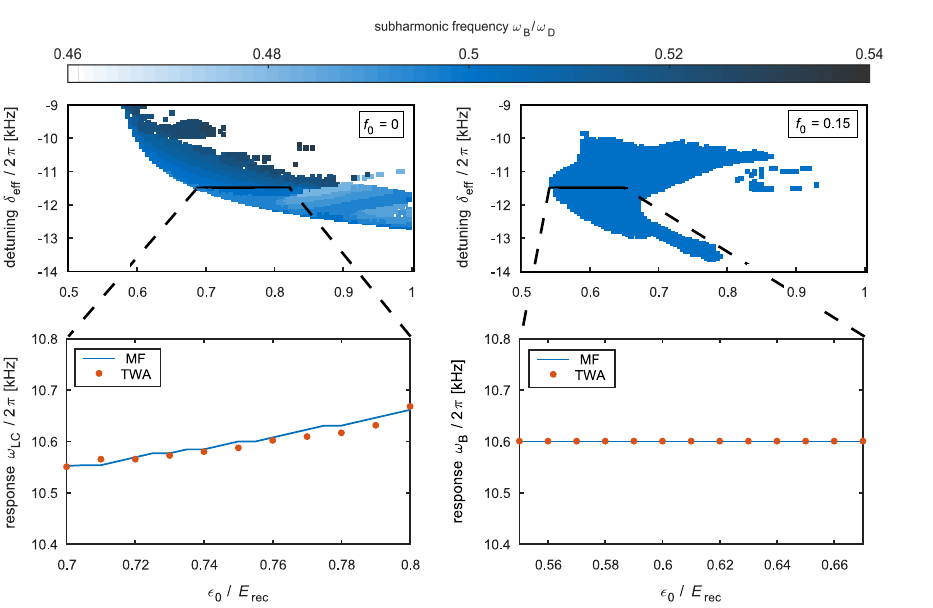}
\caption{Top: comparison between the dynamical phase diagram for the unmodulated (left) and modulated cases (right). Bottom: The results of mean-field (MF) and truncated Wigner approximation (TWA) are compared for fixed $\delta_{\mathrm{eff}}=-11.5~\mathrm{kHz}$ for a section along the horizontal lines in the top panel. To obtain better frequency resolution the driving protocol, described at the beginning of Section \ref{sec:resp}, is extended to $160~\mathrm{ms}$. The other parameters are the same as in Fig.~\ref{fig:4}.}
\label{fig:5} 
\end{figure*}
A phase diagram of the unmodulated atom-cavity system has been previously discussed in Ref.~\cite{Kessler2019}. It quantifies the long-time behavior of the cavity mode as a function of the effective detuning $\delta_{\mathrm{eff}}$ and the zero-frequency component of the pump strength $\epsilon_0$. In Fig.~\ref{fig:4}, we explore how this phase diagram changes when modulation is applied at different strengths $f_0$ between zero and $0.2$. We thereby restrict ourselves to consider only those regions, where only a single frequency component below the fixed modulation frequency $\omega_{\mathrm{D}}/2\pi = 21.2~\mathrm{kHz}$ is observed in $\mathcal{F}\{|\alpha|^2\}$. For these locations in the ($\epsilon_0$, $\delta_{\mathrm{eff}}$)-plane, we plot the value of that frequency in terms of a color scale. Regions showing no time dependance as well as regions showing chaotic and pre-chaotic dynamics with multiple spectral components appear in white.  

In the absence of periodic driving of the pump, it has been shown that the limit-cycle frequency varies weakly with the effective detuning and pump strength \cite{Kessler2019}. 
This behavior is reproduced in the phase diagram for the unmodulated case, $f_0=0$ in Fig.~\ref{fig:4}, where the response frequency $\omega_{\mathrm{B}}$ is equivalent to the limit-cycle frequency $\omega_{\mathrm{LC}}/2\pi \in~[10.2,11.2]~\mathrm{kHz}$ in our previous work \cite{Kessler2019}. 
The weak dependence of $\omega_{\mathrm{LC}}$ on the system parameters $\delta_{\mathrm{eff}}$ and $\epsilon_0$ can be attributed to the time translation invariance of the equations of motion and is expected for the CTC case \cite{Iemini2018, Lledo2019, Kessler2019}. For periodic driving near $\omega_\mathrm{D}/2\pi=2\omega_{\mathrm{LC}}/2\pi \in~[20.4,22]~\mathrm{kHz}$, the activation of a parametric resonance pushes the system to enter the DTC phase. This is highlighted in Fig.~\ref{fig:4} for sufficiently strong modulation $f_0 > 0.01$, where the subharmonic frequency is fixed at precisely $\omega_{\mathrm{B}}=\omega_{\mathrm{D}}/2$ over a large area in the ($\delta_{\mathrm{eff}},\epsilon_0$)-plane. In this case of a DTC, the emergent oscillation acquires perfect rigidity against perturbations in $\delta_{\mathrm{eff}}$ and $\epsilon_0$. Furthermore, Fig.~\ref{fig:4} shows that the size of the region, where stable DTC dynamics arises for the modulated case, can exceed that of stable CTC dynamics in the unmodulated case. The region of stable time crystal dynamics grows for increasing $f_0$ within the interval $[0.01, 0.15]$ and shrinks again for $f_0 > 0.15$. In particular, the boundary between the self-organized density wave phase and the limit cycle phase is renormalized to lower values of $\epsilon_0$ for optimal modulation.

\section{Persistence and robustness against noise}

Finally, we test the rigidity of the time crystalline response from perturbations inherent in the system. Therefore, we introduce quantum noise by employing the truncated Wigner approximation (TWA) \cite{Polkovnikov2010, Blakie2008, Carusotto2013}. Quantum dynamics is approximated within TWA by treating bosonic operators as $c$-numbers. The necessary equations of motion are then solved for an ensemble of initial states or trajectories. This procedure ensures that the initial Wigner distribution is properly sampled. In doing so, the leading-order quantum many-body effects in the dynamics are included in the semiclassical theory. For dissipative systems \cite{Carusotto2013}, the TWA also takes into account the stochastic noise $\xi$ associated with the non-unitary evolution due to photon loss in the cavity. Using TWA, we investigate the robustness of the time translation symmetry-breaking dynamics against the initial quantum noise and the stochastic noise from the dissipation \cite{Cosme2019, Kessler2019, Seibold2020}. 

The following TWA simulations are initialized by sampling the quantum noise for a condensate in the lowest momentum mode and by populating the other available but initially unoccupied cavity and atomic modes with vacuum fluctuations. We use $10^3$ trajectories to sample the initial quantum noise.
The  ensemble averaging within the TWA for the dynamics of the cavity occupation leads to a sampling of all possible realizations of the symmetry breaking response, which is quantified by the relative time phase between the external drive and the cavity mode occupation. In the case of a period-doubling DTC phase, TWA equally samples both the in-phase (0-phase shifted with respect to the modulation) and the out-of-phase ($\pi$-phase shifted) response of the cavity occupation. This causes the period-doubling inferred from the cavity mode occupation $\langle \alpha^{\dagger}(t)\alpha(t)\rangle$ to average out in TWA. To quantify time crystallinity and temporal coherence in TWA, we instead obtain the subharmonic frequency response from the two-time correlation function $\mathcal{C}(t,t_0) = \langle \alpha^{\dagger}(t)  \alpha^{\dagger}(t_0)\rangle$ at $t_0=20~\mathrm{ms}$. This quantity is insensitive to the relative time phase and the persistence of symmetry-breaking oscillations in $\mathcal{C}(t,t_0)$ is a signature of a time crystalline behavior \cite{Watanabe2015,Kessler2019}.

A comparison between the mean-field (MF) and TWA results is shown in Fig.~\ref{fig:5}. In general, the MF and TWA results yield the same qualitative behavior for the limit cycle phase both in the unmodulated and modulated cases, as seen in the lower panels of Fig.~\ref{fig:5}. Note that we extended the time window used in constructing the Fourier spectra to 160 ms in order to avoid discrete steps in $\omega_{\mathrm{B}}$ due to the finite resolution of the Fourier spectra. For the DTC phase in Fig.~\ref{fig:5}, the rigidity of the subharmonic frequency survives in the presence of inherent perturbations captured by TWA. 
Moreover, we observe that the emergent oscillations in the CTC and DTC phases persist beyond experimentally accessible timescales $t>100~\mathrm{ms}$, without any sign of decay. This suggests that the time crystalline phases discussed here could potentially persist to infinite time owing to the dissipative cavity, which counteracts driving-induced heating.

\section{Summary and discussion}\label{sec:conc}
In summary, we have discussed the dynamical stabilization of the limit-cycle phase in an experimentally relevant atom-cavity system pumped by a light field detuned to the blue side of the atomic resonance, a system that has been previously identified as a CTC \cite{Kessler2019}. In this work, we consider modulation of the pump field close to the primary parametric resonance corresponding to the characteristic frequency of the limit-cycle phase. As a result, for sufficiently large modulation strengths, period doubling dynamics emerges. Mapping the dynamical phase diagram, in combination with TWA results, demonstrates the robustness of the dynamical phase against perturbations from relevant parameters, such as the effective detuning and pump strength, and inherent sources of noise, namely the initial quantum noise and fluctuations associated with the dissipation channel. Thus, we demonstrate the formation of a DTC order from an initial CTC order. The characteristic rigidity of the DTC phase contributes to the expansion of the region of stable time crystal dynamics. Our findings points out a novel route via dynamical stabilization to experimentally realize the self-organized limit-cycle phase in an atom-cavity system with blue pump-atom detuning \cite{Kessler2019, Keeling2010, Piazza2015}, which has so far remained elusive \cite{Zupancic2019}. In the context of mean-field time crystals, the DTC phase here finds an analog in the so-called dynamical normal phase (DNP) predicted for red-detuned pump fields \cite{Chitra2015, Molignini2018}. The DNP is the mean-field counterpart of the quintessential DTC in the driven-dissipative Dicke model \cite{Gong2018,Zhu2019}. Therefore, we discover the fundamental period-doubled DTC in the mean-field limit of an atom-cavity platform with blue pump-atom detuning. 

One of the main challenges in experimentally realizing the DTC phase with red-detuned pump fields is the fact that the period doubling dynamics, wherein the system periodically switches between the two $\mathbb{Z}_2$-symmetry broken states, is only detectable from the dynamics of the relative phase between the cavity and pump fields. Unfortunately, this information is not accessible using the \textit{in situ} number of photons emitted from the cavity, typically monitored in experiments. The DTC phase in the blue-detuned pump field considered here, however, has the advantage that the subharmonic response is seen already in the dynamics of the cavity mode occupation. This allows for a direct measurement and \textit{in situ} observation of the time translation symmetry breaking dynamics.

\begin{acknowledgments}
We acknowledge support from the Deutsche Forschungsgemeinschaft via the Collaborative Research Center SFB 925 and the Hamburg Cluster of Excellence Advanced Imaging of Matter (AIM). 

\end{acknowledgments}

\bibliography{biblio}

\providecommand{\noopsort}[1]{}\providecommand{\singleletter}[1]{#1}%
\begin{thebibliography}{51}
\expandafter\ifx\csname natexlab\endcsname\relax\def\natexlab#1{#1}\fi
\expandafter\ifx\csname bibnamefont\endcsname\relax
  \def\bibnamefont#1{#1}\fi
\expandafter\ifx\csname bibfnamefont\endcsname\relax
  \def\bibfnamefont#1{#1}\fi
\expandafter\ifx\csname citenamefont\endcsname\relax
  \def\citenamefont#1{#1}\fi
\expandafter\ifx\csname url\endcsname\relax
  \def\url#1{\texttt{#1}}\fi
\expandafter\ifx\csname urlprefix\endcsname\relax\def\urlprefix{URL }\fi
\providecommand{\bibinfo}[2]{#2}
\providecommand{\eprint}[2][]{\url{#2}}

\bibitem[{\citenamefont{Wilczek}(2012)}]{Wilczek2012}
\bibinfo{author}{\bibfnamefont{F.}~\bibnamefont{Wilczek}},
  \bibinfo{journal}{Phys. Rev. Lett.} \textbf{\bibinfo{volume}{109}},
  \bibinfo{pages}{160401} (\bibinfo{year}{2012}),
  \urlprefix\url{https://link.aps.org/doi/10.1103/PhysRevLett.109.160401}.

\bibitem[{\citenamefont{Bruno}(2013)}]{Bruno2013}
\bibinfo{author}{\bibfnamefont{P.}~\bibnamefont{Bruno}},
  \bibinfo{journal}{Phys. Rev. Lett.} \textbf{\bibinfo{volume}{111}},
  \bibinfo{pages}{070402} (\bibinfo{year}{2013}),
  \urlprefix\url{https://link.aps.org/doi/10.1103/PhysRevLett.111.070402}.

\bibitem[{\citenamefont{{Nozi{\`e}res}}(2013)}]{Noz2013}
\bibinfo{author}{\bibfnamefont{P.}~\bibnamefont{{Nozi{\`e}res}}},
  \bibinfo{journal}{EPL (Europhysics Letters)} \textbf{\bibinfo{volume}{103}},
  \bibinfo{eid}{57008} (\bibinfo{year}{2013}), \eprint{1306.6229}.

\bibitem[{\citenamefont{Watanabe and Oshikawa}(2015)}]{Watanabe2015}
\bibinfo{author}{\bibfnamefont{H.}~\bibnamefont{Watanabe}} \bibnamefont{and}
  \bibinfo{author}{\bibfnamefont{M.}~\bibnamefont{Oshikawa}},
  \bibinfo{journal}{Phys. Rev. Lett.} \textbf{\bibinfo{volume}{114}},
  \bibinfo{pages}{251603} (\bibinfo{year}{2015}),
  \urlprefix\url{https://link.aps.org/doi/10.1103/PhysRevLett.114.251603}.

\bibitem[{\citenamefont{Sacha}(2015)}]{Sacha2015}
\bibinfo{author}{\bibfnamefont{K.}~\bibnamefont{Sacha}},
  \bibinfo{journal}{Phys. Rev. A} \textbf{\bibinfo{volume}{91}},
  \bibinfo{pages}{033617} (\bibinfo{year}{2015}),
  \urlprefix\url{https://link.aps.org/doi/10.1103/PhysRevA.91.033617}.

\bibitem[{\citenamefont{{Sacha} and {Zakrzewski}}(2018)}]{Sacha2018}
\bibinfo{author}{\bibfnamefont{K.}~\bibnamefont{{Sacha}}} \bibnamefont{and}
  \bibinfo{author}{\bibfnamefont{J.}~\bibnamefont{{Zakrzewski}}},
  \bibinfo{journal}{Rep. Prog. Phys.} \textbf{\bibinfo{volume}{81}},
  \bibinfo{eid}{016401} (\bibinfo{year}{2018}).

\bibitem[{\citenamefont{{Else} et~al.}(2020)\citenamefont{{Else}, {Monroe},
  {Nayak}, and {Yao}}}]{Else2019}
\bibinfo{author}{\bibfnamefont{D.~V.} \bibnamefont{{Else}}},
  \bibinfo{author}{\bibfnamefont{C.}~\bibnamefont{{Monroe}}},
  \bibinfo{author}{\bibfnamefont{C.}~\bibnamefont{{Nayak}}}, \bibnamefont{and}
  \bibinfo{author}{\bibfnamefont{N.~Y.} \bibnamefont{{Yao}}},
  \bibinfo{journal}{{Annu. Rev. Condens. Matter Phys.}}
  \textbf{\bibinfo{volume}{11}}, \bibinfo{pages}{467} (\bibinfo{year}{2020}).

\bibitem[{\citenamefont{{Khemani} et~al.}(2019)\citenamefont{{Khemani},
  {Moessner}, and {Sondhi}}}]{Khemani2019}
\bibinfo{author}{\bibfnamefont{V.}~\bibnamefont{{Khemani}}},
  \bibinfo{author}{\bibfnamefont{R.}~\bibnamefont{{Moessner}}},
  \bibnamefont{and} \bibinfo{author}{\bibfnamefont{S.~L.}
  \bibnamefont{{Sondhi}}}, \bibinfo{journal}{arXiv e-prints}
  \bibinfo{eid}{arXiv:1910.10745} (\bibinfo{year}{2019}), \eprint{1910.10745}.

\bibitem[{\citenamefont{Else et~al.}(2016)\citenamefont{Else, Bauer, and
  Nayak}}]{Else2016}
\bibinfo{author}{\bibfnamefont{D.~V.} \bibnamefont{Else}},
  \bibinfo{author}{\bibfnamefont{B.}~\bibnamefont{Bauer}}, \bibnamefont{and}
  \bibinfo{author}{\bibfnamefont{C.}~\bibnamefont{Nayak}},
  \bibinfo{journal}{Phys. Rev. Lett.} \textbf{\bibinfo{volume}{117}},
  \bibinfo{pages}{090402} (\bibinfo{year}{2016}),
  \urlprefix\url{https://link.aps.org/doi/10.1103/PhysRevLett.117.090402}.

\bibitem[{\citenamefont{Yao et~al.}(2017)\citenamefont{Yao, Potter, Potirniche,
  and Vishwanath}}]{Yao2017}
\bibinfo{author}{\bibfnamefont{N.~Y.} \bibnamefont{Yao}},
  \bibinfo{author}{\bibfnamefont{A.~C.} \bibnamefont{Potter}},
  \bibinfo{author}{\bibfnamefont{I.-D.} \bibnamefont{Potirniche}},
  \bibnamefont{and}
  \bibinfo{author}{\bibfnamefont{A.}~\bibnamefont{Vishwanath}},
  \bibinfo{journal}{Phys. Rev. Lett.} \textbf{\bibinfo{volume}{118}},
  \bibinfo{pages}{030401} (\bibinfo{year}{2017}),
  \urlprefix\url{https://link.aps.org/doi/10.1103/PhysRevLett.118.030401}.

\bibitem[{\citenamefont{{Choi} et~al.}(2017)\citenamefont{{Choi}, {Choi},
  {Landig}, {Kucsko}, {Zhou}, {Isoya}, {Jelezko}, {Onoda}, {Sumiya}, {Khemani}
  et~al.}}]{Choi2017}
\bibinfo{author}{\bibfnamefont{S.}~\bibnamefont{{Choi}}},
  \bibinfo{author}{\bibfnamefont{J.}~\bibnamefont{{Choi}}},
  \bibinfo{author}{\bibfnamefont{R.}~\bibnamefont{{Landig}}},
  \bibinfo{author}{\bibfnamefont{G.}~\bibnamefont{{Kucsko}}},
  \bibinfo{author}{\bibfnamefont{H.}~\bibnamefont{{Zhou}}},
  \bibinfo{author}{\bibfnamefont{J.}~\bibnamefont{{Isoya}}},
  \bibinfo{author}{\bibfnamefont{F.}~\bibnamefont{{Jelezko}}},
  \bibinfo{author}{\bibfnamefont{S.}~\bibnamefont{{Onoda}}},
  \bibinfo{author}{\bibfnamefont{H.}~\bibnamefont{{Sumiya}}},
  \bibinfo{author}{\bibfnamefont{V.}~\bibnamefont{{Khemani}}},
  \bibnamefont{et~al.}, \bibinfo{journal}{Nature}
  \textbf{\bibinfo{volume}{543}}, \bibinfo{pages}{221} (\bibinfo{year}{2017}).

\bibitem[{\citenamefont{{Zhang} et~al.}(2017)\citenamefont{{Zhang}, {Hess},
  {Kyprianidis}, {Becker}, {Lee}, {Smith}, {Pagano}, {Potirniche}, {Potter},
  {Vishwanath} et~al.}}]{Zhang2017}
\bibinfo{author}{\bibfnamefont{J.}~\bibnamefont{{Zhang}}},
  \bibinfo{author}{\bibfnamefont{P.~W.} \bibnamefont{{Hess}}},
  \bibinfo{author}{\bibfnamefont{A.}~\bibnamefont{{Kyprianidis}}},
  \bibinfo{author}{\bibfnamefont{P.}~\bibnamefont{{Becker}}},
  \bibinfo{author}{\bibfnamefont{A.}~\bibnamefont{{Lee}}},
  \bibinfo{author}{\bibfnamefont{J.}~\bibnamefont{{Smith}}},
  \bibinfo{author}{\bibfnamefont{G.}~\bibnamefont{{Pagano}}},
  \bibinfo{author}{\bibfnamefont{I.-D.} \bibnamefont{{Potirniche}}},
  \bibinfo{author}{\bibfnamefont{A.~C.} \bibnamefont{{Potter}}},
  \bibinfo{author}{\bibfnamefont{A.}~\bibnamefont{{Vishwanath}}},
  \bibnamefont{et~al.}, \bibinfo{journal}{Nature}
  \textbf{\bibinfo{volume}{543}}, \bibinfo{pages}{217} (\bibinfo{year}{2017}).

\bibitem[{\citenamefont{Ho et~al.}(2017)\citenamefont{Ho, Choi, Lukin, and
  Abanin}}]{Ho2017}
\bibinfo{author}{\bibfnamefont{W.~W.} \bibnamefont{Ho}},
  \bibinfo{author}{\bibfnamefont{S.}~\bibnamefont{Choi}},
  \bibinfo{author}{\bibfnamefont{M.~D.} \bibnamefont{Lukin}}, \bibnamefont{and}
  \bibinfo{author}{\bibfnamefont{D.~A.} \bibnamefont{Abanin}},
  \bibinfo{journal}{Phys. Rev. Lett.} \textbf{\bibinfo{volume}{119}},
  \bibinfo{pages}{010602} (\bibinfo{year}{2017}),
  \urlprefix\url{https://link.aps.org/doi/10.1103/PhysRevLett.119.010602}.

\bibitem[{\citenamefont{Else et~al.}(2017)\citenamefont{Else, Bauer, and
  Nayak}}]{Else2017}
\bibinfo{author}{\bibfnamefont{D.~V.} \bibnamefont{Else}},
  \bibinfo{author}{\bibfnamefont{B.}~\bibnamefont{Bauer}}, \bibnamefont{and}
  \bibinfo{author}{\bibfnamefont{C.}~\bibnamefont{Nayak}},
  \bibinfo{journal}{Phys. Rev. X} \textbf{\bibinfo{volume}{7}},
  \bibinfo{pages}{011026} (\bibinfo{year}{2017}),
  \urlprefix\url{https://link.aps.org/doi/10.1103/PhysRevX.7.011026}.

\bibitem[{\citenamefont{Rovny et~al.}(2018)\citenamefont{Rovny, Blum, and
  Barrett}}]{Rovny2018}
\bibinfo{author}{\bibfnamefont{J.}~\bibnamefont{Rovny}},
  \bibinfo{author}{\bibfnamefont{R.~L.} \bibnamefont{Blum}}, \bibnamefont{and}
  \bibinfo{author}{\bibfnamefont{S.~E.} \bibnamefont{Barrett}},
  \bibinfo{journal}{Phys. Rev. Lett.} \textbf{\bibinfo{volume}{120}},
  \bibinfo{pages}{180603} (\bibinfo{year}{2018}),
  \urlprefix\url{https://link.aps.org/doi/10.1103/PhysRevLett.120.180603}.

\bibitem[{\citenamefont{{Smits} et~al.}(2018)\citenamefont{{Smits}, {Liao},
  {Stoof}, and {van der Straten}}}]{Smits2018}
\bibinfo{author}{\bibfnamefont{J.}~\bibnamefont{{Smits}}},
  \bibinfo{author}{\bibfnamefont{L.}~\bibnamefont{{Liao}}},
  \bibinfo{author}{\bibfnamefont{H.~T.~C.} \bibnamefont{{Stoof}}},
  \bibnamefont{and} \bibinfo{author}{\bibfnamefont{P.}~\bibnamefont{{van der
  Straten}}}, \bibinfo{journal}{Phys. Rev. Lett.}
  \textbf{\bibinfo{volume}{121}}, \bibinfo{pages}{185301}
  (\bibinfo{year}{2018}),
  \urlprefix\url{https://link.aps.org/doi/10.1103/PhysRevLett.121.185301}.

\bibitem[{\citenamefont{Giergiel et~al.}(2018)\citenamefont{Giergiel, Kosior,
  Hannaford, and Sacha}}]{Giergiel2018}
\bibinfo{author}{\bibfnamefont{K.}~\bibnamefont{Giergiel}},
  \bibinfo{author}{\bibfnamefont{A.}~\bibnamefont{Kosior}},
  \bibinfo{author}{\bibfnamefont{P.}~\bibnamefont{Hannaford}},
  \bibnamefont{and} \bibinfo{author}{\bibfnamefont{K.}~\bibnamefont{Sacha}},
  \bibinfo{journal}{Phys. Rev. A} \textbf{\bibinfo{volume}{98}},
  \bibinfo{pages}{013613} (\bibinfo{year}{2018}),
  \urlprefix\url{https://link.aps.org/doi/10.1103/PhysRevA.98.013613}.

\bibitem[{\citenamefont{Giergiel et~al.}(2019)\citenamefont{Giergiel,
  Kuro\ifmmode~\acute{s}\else \'{s}\fi{}, and Sacha}}]{Giergiel2019}
\bibinfo{author}{\bibfnamefont{K.}~\bibnamefont{Giergiel}},
  \bibinfo{author}{\bibfnamefont{A.}~\bibnamefont{Kuro\ifmmode~\acute{s}\else
  \'{s}\fi{}}}, \bibnamefont{and}
  \bibinfo{author}{\bibfnamefont{K.}~\bibnamefont{Sacha}},
  \bibinfo{journal}{Phys. Rev. B} \textbf{\bibinfo{volume}{99}},
  \bibinfo{pages}{220303(R)} (\bibinfo{year}{2019}),
  \urlprefix\url{https://link.aps.org/doi/10.1103/PhysRevB.99.220303}.

\bibitem[{\citenamefont{{Pizzi} et~al.}(2019)\citenamefont{{Pizzi}, {Knolle},
  and {Nunnenkamp}}}]{Pizzi2019}
\bibinfo{author}{\bibfnamefont{A.}~\bibnamefont{{Pizzi}}},
  \bibinfo{author}{\bibfnamefont{J.}~\bibnamefont{{Knolle}}}, \bibnamefont{and}
  \bibinfo{author}{\bibfnamefont{A.}~\bibnamefont{{Nunnenkamp}}},
  \bibinfo{journal}{Phys. Rev. Lett.} \textbf{\bibinfo{volume}{123}},
  \bibinfo{pages}{150601} (\bibinfo{year}{2019}),
  \urlprefix\url{https://link.aps.org/doi/10.1103/PhysRevLett.123.150601}.

\bibitem[{\citenamefont{Autti et~al.}(2018)\citenamefont{Autti, Eltsov, and
  Volovik}}]{Autti2018}
\bibinfo{author}{\bibfnamefont{S.}~\bibnamefont{Autti}},
  \bibinfo{author}{\bibfnamefont{V.~B.} \bibnamefont{Eltsov}},
  \bibnamefont{and} \bibinfo{author}{\bibfnamefont{G.~E.}
  \bibnamefont{Volovik}}, \bibinfo{journal}{Phys. Rev. Lett.}
  \textbf{\bibinfo{volume}{120}}, \bibinfo{pages}{215301}
  (\bibinfo{year}{2018}),
  \urlprefix\url{https://link.aps.org/doi/10.1103/PhysRevLett.120.215301}.

\bibitem[{\citenamefont{Gong et~al.}(2018)\citenamefont{Gong, Hamazaki, and
  Ueda}}]{Gong2018}
\bibinfo{author}{\bibfnamefont{Z.}~\bibnamefont{Gong}},
  \bibinfo{author}{\bibfnamefont{R.}~\bibnamefont{Hamazaki}}, \bibnamefont{and}
  \bibinfo{author}{\bibfnamefont{M.}~\bibnamefont{Ueda}},
  \bibinfo{journal}{Phys. Rev. Lett.} \textbf{\bibinfo{volume}{120}},
  \bibinfo{pages}{040404} (\bibinfo{year}{2018}),
  \urlprefix\url{https://link.aps.org/doi/10.1103/PhysRevLett.120.040404}.

\bibitem[{\citenamefont{{Gambetta} et~al.}(2019)\citenamefont{{Gambetta},
  {Carollo}, {Marcuzzi}, {Garrahan}, and {Lesanovsky}}}]{Gambetta2018}
\bibinfo{author}{\bibfnamefont{F.~M.} \bibnamefont{{Gambetta}}},
  \bibinfo{author}{\bibfnamefont{F.}~\bibnamefont{{Carollo}}},
  \bibinfo{author}{\bibfnamefont{M.}~\bibnamefont{{Marcuzzi}}},
  \bibinfo{author}{\bibfnamefont{J.~P.} \bibnamefont{{Garrahan}}},
  \bibnamefont{and}
  \bibinfo{author}{\bibfnamefont{I.}~\bibnamefont{{Lesanovsky}}},
  \bibinfo{journal}{Phys. Rev. Lett.} \textbf{\bibinfo{volume}{122}},
  \bibinfo{pages}{015701} (\bibinfo{year}{2019}),
  \urlprefix\url{https://link.aps.org/doi/10.1103/PhysRevLett.122.015701}.

\bibitem[{\citenamefont{Iemini et~al.}(2018)\citenamefont{Iemini, Russomanno,
  Keeling, Schir\`o, Dalmonte, and Fazio}}]{Iemini2018}
\bibinfo{author}{\bibfnamefont{F.}~\bibnamefont{Iemini}},
  \bibinfo{author}{\bibfnamefont{A.}~\bibnamefont{Russomanno}},
  \bibinfo{author}{\bibfnamefont{J.}~\bibnamefont{Keeling}},
  \bibinfo{author}{\bibfnamefont{M.}~\bibnamefont{Schir\`o}},
  \bibinfo{author}{\bibfnamefont{M.}~\bibnamefont{Dalmonte}}, \bibnamefont{and}
  \bibinfo{author}{\bibfnamefont{R.}~\bibnamefont{Fazio}},
  \bibinfo{journal}{Phys. Rev. Lett.} \textbf{\bibinfo{volume}{121}},
  \bibinfo{pages}{035301} (\bibinfo{year}{2018}),
  \urlprefix\url{https://link.aps.org/doi/10.1103/PhysRevLett.121.035301}.

\bibitem[{\citenamefont{{Tucker} et~al.}(2018)\citenamefont{{Tucker}, {Zhu},
  {Lewis-Swan}, {Marino}, {Jimenez}, {Restrepo}, and {Rey}}}]{Tucker2018}
\bibinfo{author}{\bibfnamefont{K.}~\bibnamefont{{Tucker}}},
  \bibinfo{author}{\bibfnamefont{B.}~\bibnamefont{{Zhu}}},
  \bibinfo{author}{\bibfnamefont{R.~J.} \bibnamefont{{Lewis-Swan}}},
  \bibinfo{author}{\bibfnamefont{J.}~\bibnamefont{{Marino}}},
  \bibinfo{author}{\bibfnamefont{F.}~\bibnamefont{{Jimenez}}},
  \bibinfo{author}{\bibfnamefont{J.~G.} \bibnamefont{{Restrepo}}},
  \bibnamefont{and} \bibinfo{author}{\bibfnamefont{A.~M.} \bibnamefont{{Rey}}},
  \bibinfo{journal}{New J. Phys.} \textbf{\bibinfo{volume}{20}},
  \bibinfo{eid}{123003} (\bibinfo{year}{2018}).

\bibitem[{\citenamefont{{Heugel} et~al.}(2019)\citenamefont{{Heugel}, {Oscity},
  {Eichler}, {Zilberberg}, and {Chitra}}}]{Heugel2019}
\bibinfo{author}{\bibfnamefont{T.~L.} \bibnamefont{{Heugel}}},
  \bibinfo{author}{\bibfnamefont{M.}~\bibnamefont{{Oscity}}},
  \bibinfo{author}{\bibfnamefont{A.}~\bibnamefont{{Eichler}}},
  \bibinfo{author}{\bibfnamefont{O.}~\bibnamefont{{Zilberberg}}},
  \bibnamefont{and} \bibinfo{author}{\bibfnamefont{R.}~\bibnamefont{{Chitra}}},
  \bibinfo{journal}{Phys. Rev. Lett.} \textbf{\bibinfo{volume}{123}},
  \bibinfo{pages}{124301} (\bibinfo{year}{2019}),
  \urlprefix\url{https://link.aps.org/doi/10.1103/PhysRevLett.123.124301}.

\bibitem[{\citenamefont{Lazarides et~al.}(2020)\citenamefont{Lazarides, Roy,
  Piazza, and Moessner}}]{Lazarides2019}
\bibinfo{author}{\bibfnamefont{A.}~\bibnamefont{Lazarides}},
  \bibinfo{author}{\bibfnamefont{S.}~\bibnamefont{Roy}},
  \bibinfo{author}{\bibfnamefont{F.}~\bibnamefont{Piazza}}, \bibnamefont{and}
  \bibinfo{author}{\bibfnamefont{R.}~\bibnamefont{Moessner}},
  \bibinfo{journal}{Phys. Rev. Research} \textbf{\bibinfo{volume}{2}},
  \bibinfo{pages}{022002} (\bibinfo{year}{2020}),
  \urlprefix\url{https://link.aps.org/doi/10.1103/PhysRevResearch.2.022002}.

\bibitem[{\citenamefont{{Zhu} et~al.}(2019)\citenamefont{{Zhu}, {Marino},
  {Yao}, {Lukin}, and {Demler}}}]{Zhu2019}
\bibinfo{author}{\bibfnamefont{B.}~\bibnamefont{{Zhu}}},
  \bibinfo{author}{\bibfnamefont{J.}~\bibnamefont{{Marino}}},
  \bibinfo{author}{\bibfnamefont{N.~Y.} \bibnamefont{{Yao}}},
  \bibinfo{author}{\bibfnamefont{M.~D.} \bibnamefont{{Lukin}}},
  \bibnamefont{and} \bibinfo{author}{\bibfnamefont{E.~A.}
  \bibnamefont{{Demler}}}, \bibinfo{journal}{New J. Phys.}
  \textbf{\bibinfo{volume}{21}}, \bibinfo{pages}{073028}
  (\bibinfo{year}{2019}),
  \urlprefix\url{https://doi.org/10.1088%2F1367-2630%2Fab2afe}.

\bibitem[{\citenamefont{{Lled{\'o}} et~al.}(2019)\citenamefont{{Lled{\'o}},
  {Mavrogordatos}, and {Szyma{\'n}ska}}}]{Lledo2019}
\bibinfo{author}{\bibfnamefont{C.}~\bibnamefont{{Lled{\'o}}}},
  \bibinfo{author}{\bibfnamefont{T.~K.} \bibnamefont{{Mavrogordatos}}},
  \bibnamefont{and} \bibinfo{author}{\bibfnamefont{M.~H.}
  \bibnamefont{{Szyma{\'n}ska}}}, \bibinfo{journal}{Phys. Rev. B}
  \textbf{\bibinfo{volume}{100}}, \bibinfo{pages}{054303}
  (\bibinfo{year}{2019}),
  \urlprefix\url{https://link.aps.org/doi/10.1103/PhysRevB.100.054303}.

\bibitem[{\citenamefont{{Buca} et~al.}(2019)\citenamefont{{Buca}, {Tindall},
  and {Jaksch}}}]{Buca2019}
\bibinfo{author}{\bibfnamefont{B.}~\bibnamefont{{Buca}}},
  \bibinfo{author}{\bibfnamefont{J.}~\bibnamefont{{Tindall}}},
  \bibnamefont{and} \bibinfo{author}{\bibfnamefont{D.}~\bibnamefont{{Jaksch}}},
  \bibinfo{journal}{Nat. Commun.} \textbf{\bibinfo{volume}{10}},
  \bibinfo{pages}{1730} (\bibinfo{year}{2019}).

\bibitem[{\citenamefont{{Ke{\ss}ler} et~al.}(2019)\citenamefont{{Ke{\ss}ler},
  {Cosme}, {Hemmerling}, {Mathey}, and {Hemmerich}}}]{Kessler2019}
\bibinfo{author}{\bibfnamefont{H.}~\bibnamefont{{Ke{\ss}ler}}},
  \bibinfo{author}{\bibfnamefont{J.~G.} \bibnamefont{{Cosme}}},
  \bibinfo{author}{\bibfnamefont{M.}~\bibnamefont{{Hemmerling}}},
  \bibinfo{author}{\bibfnamefont{L.}~\bibnamefont{{Mathey}}}, \bibnamefont{and}
  \bibinfo{author}{\bibfnamefont{A.}~\bibnamefont{{Hemmerich}}},
  \bibinfo{journal}{Phys. Rev. A} \textbf{\bibinfo{volume}{99}},
  \bibinfo{pages}{053605} (\bibinfo{year}{2019}).

\bibitem[{\citenamefont{Cosme et~al.}(2019)\citenamefont{Cosme, Skulte, and
  Mathey}}]{Cosme2019}
\bibinfo{author}{\bibfnamefont{J.~G.} \bibnamefont{Cosme}},
  \bibinfo{author}{\bibfnamefont{J.}~\bibnamefont{Skulte}}, \bibnamefont{and}
  \bibinfo{author}{\bibfnamefont{L.}~\bibnamefont{Mathey}},
  \bibinfo{journal}{Phys. Rev. A} \textbf{\bibinfo{volume}{100}},
  \bibinfo{pages}{053615} (\bibinfo{year}{2019}),
  \urlprefix\url{https://link.aps.org/doi/10.1103/PhysRevA.100.053615}.

\bibitem[{\citenamefont{Seibold et~al.}(2020)\citenamefont{Seibold, Rota, and
  Savona}}]{Seibold2020}
\bibinfo{author}{\bibfnamefont{K.}~\bibnamefont{Seibold}},
  \bibinfo{author}{\bibfnamefont{R.}~\bibnamefont{Rota}}, \bibnamefont{and}
  \bibinfo{author}{\bibfnamefont{V.}~\bibnamefont{Savona}},
  \bibinfo{journal}{Phys. Rev. A} \textbf{\bibinfo{volume}{101}},
  \bibinfo{pages}{033839} (\bibinfo{year}{2020}),
  \urlprefix\url{https://link.aps.org/doi/10.1103/PhysRevA.101.033839}.

\bibitem[{\citenamefont{Yao et~al.}(2020)\citenamefont{Yao, Nayak, Balents, and
  Zaletel}}]{Yao2020}
\bibinfo{author}{\bibfnamefont{N.~Y.} \bibnamefont{Yao}},
  \bibinfo{author}{\bibfnamefont{C.}~\bibnamefont{Nayak}},
  \bibinfo{author}{\bibfnamefont{L.}~\bibnamefont{Balents}}, \bibnamefont{and}
  \bibinfo{author}{\bibfnamefont{M.~P.} \bibnamefont{Zaletel}},
  \bibinfo{journal}{Nat. Phys.} \textbf{\bibinfo{volume}{16}},
  \bibinfo{pages}{438} (\bibinfo{year}{2020}),
  \urlprefix\url{https://www.nature.com/articles/s41567-019-0782-3}.

\bibitem[{\citenamefont{{Lled{\'o}} and {Szyma{\'n}ska}}(2020)}]{Lledo2020}
\bibinfo{author}{\bibfnamefont{C.}~\bibnamefont{{Lled{\'o}}}} \bibnamefont{and}
  \bibinfo{author}{\bibfnamefont{M.~H.} \bibnamefont{{Szyma{\'n}ska}}},
  \bibinfo{journal}{arXiv e-prints} \bibinfo{eid}{arXiv:2004.02855}
  (\bibinfo{year}{2020}), \eprint{2004.02855}.

\bibitem[{\citenamefont{Keeling et~al.}(2010)\citenamefont{Keeling, Bhaseen,
  and Simons}}]{Keeling2010}
\bibinfo{author}{\bibfnamefont{J.}~\bibnamefont{Keeling}},
  \bibinfo{author}{\bibfnamefont{M.~J.} \bibnamefont{Bhaseen}},
  \bibnamefont{and} \bibinfo{author}{\bibfnamefont{B.~D.}
  \bibnamefont{Simons}}, \bibinfo{journal}{Phys. Rev. Lett.}
  \textbf{\bibinfo{volume}{105}}, \bibinfo{pages}{043001}
  (\bibinfo{year}{2010}),
  \urlprefix\url{https://link.aps.org/doi/10.1103/PhysRevLett.105.043001}.

\bibitem[{\citenamefont{Piazza and Ritsch}(2015)}]{Piazza2015}
\bibinfo{author}{\bibfnamefont{F.}~\bibnamefont{Piazza}} \bibnamefont{and}
  \bibinfo{author}{\bibfnamefont{H.}~\bibnamefont{Ritsch}},
  \bibinfo{journal}{Phys. Rev. Lett.} \textbf{\bibinfo{volume}{115}},
  \bibinfo{pages}{163601} (\bibinfo{year}{2015}),
  \urlprefix\url{https://link.aps.org/doi/10.1103/PhysRevLett.115.163601}.

\bibitem[{\citenamefont{Ritsch et~al.}(2013)\citenamefont{Ritsch, Domokos,
  Brennecke, and Esslinger}}]{Ritsch2013}
\bibinfo{author}{\bibfnamefont{H.}~\bibnamefont{Ritsch}},
  \bibinfo{author}{\bibfnamefont{P.}~\bibnamefont{Domokos}},
  \bibinfo{author}{\bibfnamefont{F.}~\bibnamefont{Brennecke}},
  \bibnamefont{and}
  \bibinfo{author}{\bibfnamefont{T.}~\bibnamefont{Esslinger}},
  \bibinfo{journal}{Rev. Mod. Phys.} \textbf{\bibinfo{volume}{85}},
  \bibinfo{pages}{553} (\bibinfo{year}{2013}),
  \urlprefix\url{https://link.aps.org/doi/10.1103/RevModPhys.85.553}.

\bibitem[{\citenamefont{Zupancic et~al.}(2019)\citenamefont{Zupancic, Dreon,
  Li, Baumg\"artner, Morales, Zheng, Cooper, Esslinger, and
  Donner}}]{Zupancic2019}
\bibinfo{author}{\bibfnamefont{P.}~\bibnamefont{Zupancic}},
  \bibinfo{author}{\bibfnamefont{D.}~\bibnamefont{Dreon}},
  \bibinfo{author}{\bibfnamefont{X.}~\bibnamefont{Li}},
  \bibinfo{author}{\bibfnamefont{A.}~\bibnamefont{Baumg\"artner}},
  \bibinfo{author}{\bibfnamefont{A.}~\bibnamefont{Morales}},
  \bibinfo{author}{\bibfnamefont{W.}~\bibnamefont{Zheng}},
  \bibinfo{author}{\bibfnamefont{N.~R.} \bibnamefont{Cooper}},
  \bibinfo{author}{\bibfnamefont{T.}~\bibnamefont{Esslinger}},
  \bibnamefont{and} \bibinfo{author}{\bibfnamefont{T.}~\bibnamefont{Donner}},
  \bibinfo{journal}{Phys. Rev. Lett.} \textbf{\bibinfo{volume}{123}},
  \bibinfo{pages}{233601} (\bibinfo{year}{2019}),
  \urlprefix\url{https://link.aps.org/doi/10.1103/PhysRevLett.123.233601}.

\bibitem[{\citenamefont{Domokos and Ritsch}(2002)}]{Domokos2002}
\bibinfo{author}{\bibfnamefont{P.}~\bibnamefont{Domokos}} \bibnamefont{and}
  \bibinfo{author}{\bibfnamefont{H.}~\bibnamefont{Ritsch}},
  \bibinfo{journal}{Phys. Rev. Lett.} \textbf{\bibinfo{volume}{89}},
  \bibinfo{pages}{253003} (\bibinfo{year}{2002}),
  \urlprefix\url{https://link.aps.org/doi/10.1103/PhysRevLett.89.253003}.

\bibitem[{\citenamefont{{Baumann} et~al.}(2010)\citenamefont{{Baumann},
  {Guerlin}, {Brennecke}, and {Esslinger}}}]{Baumann2010}
\bibinfo{author}{\bibfnamefont{K.}~\bibnamefont{{Baumann}}},
  \bibinfo{author}{\bibfnamefont{C.}~\bibnamefont{{Guerlin}}},
  \bibinfo{author}{\bibfnamefont{F.}~\bibnamefont{{Brennecke}}},
  \bibnamefont{and}
  \bibinfo{author}{\bibfnamefont{T.}~\bibnamefont{{Esslinger}}},
  \bibinfo{journal}{Nature} \textbf{\bibinfo{volume}{464}},
  \bibinfo{pages}{1301} (\bibinfo{year}{2010}).

\bibitem[{\citenamefont{{Klinder} et~al.}(2015)\citenamefont{{Klinder},
  {Ke{\ss}ler}, {Wolke}, {Mathey}, and {Hemmerich}}}]{Klinder2015}
\bibinfo{author}{\bibfnamefont{J.}~\bibnamefont{{Klinder}}},
  \bibinfo{author}{\bibfnamefont{H.}~\bibnamefont{{Ke{\ss}ler}}},
  \bibinfo{author}{\bibfnamefont{M.}~\bibnamefont{{Wolke}}},
  \bibinfo{author}{\bibfnamefont{L.}~\bibnamefont{{Mathey}}}, \bibnamefont{and}
  \bibinfo{author}{\bibfnamefont{A.}~\bibnamefont{{Hemmerich}}},
  \bibinfo{journal}{Proc. Natl. Acad. Sci. USA} \textbf{\bibinfo{volume}{112}},
  \bibinfo{pages}{3290} (\bibinfo{year}{2015}).

\bibitem[{\citenamefont{Lin et~al.}(2019)\citenamefont{Lin, Molignini, Lode,
  and Chitra}}]{Lin2019}
\bibinfo{author}{\bibfnamefont{R.}~\bibnamefont{Lin}},
  \bibinfo{author}{\bibfnamefont{P.}~\bibnamefont{Molignini}},
  \bibinfo{author}{\bibfnamefont{A.~U.~J.} \bibnamefont{Lode}},
  \bibnamefont{and} \bibinfo{author}{\bibfnamefont{R.}~\bibnamefont{Chitra}},
  \bibinfo{journal}{arXiv e-prints} \bibinfo{eid}{arXiv:1910.01143}
  (\bibinfo{year}{2019}), \eprint{1910.01143}.

\bibitem[{\citenamefont{Klinder et~al.}(2016)\citenamefont{Klinder, Ke{\ss}ler,
  Georges, Vargas, and Hemmerich}}]{Klinder2016}
\bibinfo{author}{\bibfnamefont{J.}~\bibnamefont{Klinder}},
  \bibinfo{author}{\bibfnamefont{H.}~\bibnamefont{Ke{\ss}ler}},
  \bibinfo{author}{\bibfnamefont{C.}~\bibnamefont{Georges}},
  \bibinfo{author}{\bibfnamefont{J.}~\bibnamefont{Vargas}}, \bibnamefont{and}
  \bibinfo{author}{\bibfnamefont{A.}~\bibnamefont{Hemmerich}},
  \bibinfo{journal}{Applied Physics B} \textbf{\bibinfo{volume}{122}},
  \bibinfo{pages}{299} (\bibinfo{year}{2016}).

\bibitem[{\citenamefont{{Georges} et~al.}(2018)\citenamefont{{Georges},
  {Cosme}, {Mathey}, and {Hemmerich}}}]{Georges2018}
\bibinfo{author}{\bibfnamefont{C.}~\bibnamefont{{Georges}}},
  \bibinfo{author}{\bibfnamefont{J.~G.} \bibnamefont{{Cosme}}},
  \bibinfo{author}{\bibfnamefont{L.}~\bibnamefont{{Mathey}}}, \bibnamefont{and}
  \bibinfo{author}{\bibfnamefont{A.}~\bibnamefont{{Hemmerich}}},
  \bibinfo{journal}{Phys. Rev. Lett.} \textbf{\bibinfo{volume}{121}},
  \bibinfo{pages}{220405} (\bibinfo{year}{2018}),
  \urlprefix\url{https://link.aps.org/doi/10.1103/PhysRevLett.121.220405}.

\bibitem[{\citenamefont{{Cosme} et~al.}(2018)\citenamefont{{Cosme}, {Georges},
  {Hemmerich}, and {Mathey}}}]{Cosme2018}
\bibinfo{author}{\bibfnamefont{J.~G.} \bibnamefont{{Cosme}}},
  \bibinfo{author}{\bibfnamefont{C.}~\bibnamefont{{Georges}}},
  \bibinfo{author}{\bibfnamefont{A.}~\bibnamefont{{Hemmerich}}},
  \bibnamefont{and} \bibinfo{author}{\bibfnamefont{L.}~\bibnamefont{{Mathey}}},
  \bibinfo{journal}{Phys. Rev. Lett.} \textbf{\bibinfo{volume}{121}},
  \bibinfo{pages}{153001} (\bibinfo{year}{2018}),
  \urlprefix\url{https://link.aps.org/doi/10.1103/PhysRevLett.121.153001}.

\bibitem[{\citenamefont{Chitra and Zilberberg}(2015)}]{Chitra2015}
\bibinfo{author}{\bibfnamefont{R.}~\bibnamefont{Chitra}} \bibnamefont{and}
  \bibinfo{author}{\bibfnamefont{O.}~\bibnamefont{Zilberberg}},
  \bibinfo{journal}{Phys. Rev. A} \textbf{\bibinfo{volume}{92}},
  \bibinfo{pages}{023815} (\bibinfo{year}{2015}),
  \urlprefix\url{https://link.aps.org/doi/10.1103/PhysRevA.92.023815}.

\bibitem[{\citenamefont{{Molignini} et~al.}(2018)\citenamefont{{Molignini},
  {Papariello}, {Lode}, and {Chitra}}}]{Molignini2018}
\bibinfo{author}{\bibfnamefont{P.}~\bibnamefont{{Molignini}}},
  \bibinfo{author}{\bibfnamefont{L.}~\bibnamefont{{Papariello}}},
  \bibinfo{author}{\bibfnamefont{A.~U.~J.} \bibnamefont{{Lode}}},
  \bibnamefont{and} \bibinfo{author}{\bibfnamefont{R.}~\bibnamefont{{Chitra}}},
  \bibinfo{journal}{Phys. Rev. A} \textbf{\bibinfo{volume}{98}},
  \bibinfo{pages}{053620} (\bibinfo{year}{2018}),
  \urlprefix\url{https://link.aps.org/doi/10.1103/PhysRevA.98.053620}.

\bibitem[{\citenamefont{{Georges} et~al.}(2020)\citenamefont{{Georges},
  {Cosme}, {Ke{\ss}ler}, {Mathey}, and {Hemmerich}}}]{Georges2020}
\bibinfo{author}{\bibfnamefont{C.}~\bibnamefont{{Georges}}},
  \bibinfo{author}{\bibfnamefont{J.~G.} \bibnamefont{{Cosme}}},
  \bibinfo{author}{\bibfnamefont{H.}~\bibnamefont{{Ke{\ss}ler}}},
  \bibinfo{author}{\bibfnamefont{L.}~\bibnamefont{{Mathey}}}, \bibnamefont{and}
  \bibinfo{author}{\bibfnamefont{A.}~\bibnamefont{{Hemmerich}}},
  \bibinfo{journal}{arXiv e-prints} \bibinfo{eid}{arXiv:2003.14135}
  (\bibinfo{year}{2020}), \eprint{2003.14135}.

\bibitem[{\citenamefont{Polkovnikov}(2010)}]{Polkovnikov2010}
\bibinfo{author}{\bibfnamefont{A.}~\bibnamefont{Polkovnikov}},
  \bibinfo{journal}{Ann. Phys.} \textbf{\bibinfo{volume}{325}},
  \bibinfo{pages}{1790} (\bibinfo{year}{2010}), ISSN \bibinfo{issn}{0003-4916},
  \urlprefix\url{http://www.sciencedirect.com/science/article/pii/S0003491610000382}.

\bibitem[{\citenamefont{Blakie et~al.}(2008)\citenamefont{Blakie, Bradley,
  Davis, Ballagh, and Gardiner}}]{Blakie2008}
\bibinfo{author}{\bibfnamefont{P.~B.} \bibnamefont{Blakie}},
  \bibinfo{author}{\bibfnamefont{A.~S.} \bibnamefont{Bradley}},
  \bibinfo{author}{\bibfnamefont{M.~J.} \bibnamefont{Davis}},
  \bibinfo{author}{\bibfnamefont{R.~J.} \bibnamefont{Ballagh}},
  \bibnamefont{and} \bibinfo{author}{\bibfnamefont{C.~W.}
  \bibnamefont{Gardiner}}, \bibinfo{journal}{Adv. Phys.}
  \textbf{\bibinfo{volume}{57}}, \bibinfo{pages}{363} (\bibinfo{year}{2008}).

\bibitem[{\citenamefont{Carusotto and Ciuti}(2013)}]{Carusotto2013}
\bibinfo{author}{\bibfnamefont{I.}~\bibnamefont{Carusotto}} \bibnamefont{and}
  \bibinfo{author}{\bibfnamefont{C.}~\bibnamefont{Ciuti}},
  \bibinfo{journal}{Rev. Mod. Phys.} \textbf{\bibinfo{volume}{85}},
  \bibinfo{pages}{299} (\bibinfo{year}{2013}),
  \urlprefix\url{https://link.aps.org/doi/10.1103/RevModPhys.85.299}.

\end{thebibliography}

\end{document}